\documentclass[twocolumn]{aastex61}

\usepackage[normalem]{ulem}

\def\lesssim{\mathrel{\hbox{\rlap{\hbox{\lower4pt\hbox{$\sim$}}}\hbox{$<$}}}}
\def\gtrsim{\mathrel{\hbox{\rlap{\hbox{\lower4pt\hbox{$\sim$}}}\hbox{$>$}}}}

\def \aj {AJ}
\def \mnras {MNRAS}
\def \apj {ApJ}
\def \apjs {ApJS}
\def \apjl {ApJL}
\def \aap {A\&A}

\begin{document}

\shorttitle{Morphological segregation in the surroundings of cosmic voids}
\shortauthors{Ricciardelli et al.}

\title{Morphological segregation in the surroundings of cosmic voids}

\correspondingauthor{Elena Ricciardelli}
\email{elena.ricciardelli@epfl.ch}

\author{Elena Ricciardelli}
\affil{Laboratoire d'Astrophysique, \'Ecole Polytechnique F\'ed\'erale de Lausanne (EPFL), 1290 Sauverny, Switzerland}

\author{Antonio Cava}
\affil{Observatoire de Gen{\`e}ve, Universit{\'e} de Gen{\`e}ve, 51 Ch. des Maillettes, 1290 Versoix, Switzerland}

\author{Jesus Varela}
\affil{Centro de Estudios de F\'{\i}sica del Cosmos de Arag\'on (CEFCA), Plaza San Juan 1, 44001 Teruel, Spain}

\author{Amelie Tamone}
\affil{Laboratoire d'Astrophysique, \'Ecole Polytechnique F\'ed\'erale de Lausanne (EPFL), 1290 Sauverny, Switzerland}

\begin{abstract}

We explore the morphology of galaxies living in the proximity of cosmic voids, using  a sample of voids identified in the Sloan Digital Sky Survey Data Release 7.
At all stellar masses, void galaxies exhibit morphologies of a later type than galaxies in a control sample, that represent galaxies in an average density environment. We interpret this trend as a pure environmental effect, independent  of the mass bias, due to a slower galaxy build-up in the rarefied regions of voids. 
We confirm previous findings about a clear segregation in galaxy morphology, with galaxies of a later type being found at smaller void-centric distances with respect to the early-type galaxies.
We also show, for the first time, that the radius of the void has an impact on the evolutionary history of the galaxies that live within it or in its surroundings.
In fact, an enhanced fraction of late-type galaxies is found in the proximity of voids larger than  the median void radius. Likewise, an excess of early-type galaxies is observed within or around voids of smaller size.
A significant difference in galaxy properties in voids of different sizes is observed up to $2\,R_{void}$, that we define as the region of influence of voids.
The significance of this difference is greater than $3\,\sigma$ for all the volume-complete samples considered  here.  The fraction of star forming galaxies shows the same behaviour as the late-type galaxies, but no significant difference in stellar mass is observed in the proximity of voids of different sizes. 

\end{abstract}
 
\keywords{ cosmology: observations --- galaxies: evolution --- large-scale structure of universe  }

\section{Introduction}\label{intro}

Cosmic voids are large underdense regions that constitute one of the most prominent aspect of the Cosmic Web. 
Within such rarefied regions, a rich infrastructure made of tenuous filaments has been found in both numerical and observational studies of voids 
\citep{Aragon13, Beygu13, Rieder13}. These filamentary structures, also called "tendrils" \citep{Alpaslan14}, are believed to be the favorite sites of galaxy formation.

The pristine environment of cosmic voids provides an ideal laboratory to study  galaxy evolution as a result of nature only, in the absence of nurture. As revealed by observational
studies of statistical samples of voids, galaxies in voids are bluer,
have higher specific star formation rates and are of later types than
galaxies living in regions at average density \citep{Rojas04, Rojas05, Patiri06, vonbenda08, Hoyle12, Kreckel12, Moorman16}.  
Part of this trend is due to an increased proportion of low-mass
galaxies in low-density regions. However, in \citet{Ricciardelli14b},
it has been shown that, at fixed stellar mass, 
the number of star forming galaxies in voids is still higher than in a
control sample, thus a residual environmental effect, beside that on
stellar mass,  is present. 

Voids can also shed light into the understanding of the effect of the
large-scale  environment on galaxy
evolution.  
Although the effect of the small-scale environment, namely
the local density, on the evolution of galaxies has been largely
addressed (e.g. \citealt{Dressler80, Kauffmann04, Bamford09}), the
extent to which the large scale structures influence the build-up of galaxies 
is much less understood. 
We can expect that the different dynamical states  of the distinct Cosmic Web components have some effect on the constituent galaxies \citep{Aragon10}.  On Megaparsec scales, the dynamics of the Cosmic Web drives matter out of the  voids, into walls and filaments before it finally gets accreted on to dark matter haloes.
Studies of filamentary environment indicate that filaments host an enhanced fraction of star forming galaxies with respect to the field \citep{Darvish14} and show a segregation in galaxy properties, with the galaxies lying closer to filaments being more massive and less star-forming \citep{Malavasi17, Kuutma17}. Furthermore, there are now plenty of evidences of the instrumental importance of the large-scale structures in advecting angular momentum on to galaxies, a fact that can be interpreted as 
 a consequence of the tidal shear produced by the neighboring primordial matter distribution \citep{Pichon11, Codis15}. 
 
In this paper, we focus on the effect of the low density environment of
voids on the morphology of void galaxies. The morphology of a galaxy is an indicator of
its current internal structure and kinematics, which in turn are a
result of the galaxy€™'s evolutionary history. 
It is widely accepted that the morphology of galaxies has a strong dependence on the local environment in which galaxies live, but at which extent this result can be extended to the extreme low density of voids, is not clear.
The structure of the paper is as follows. In Section \ref{section:data} we describe the sample used for the analysis, in Section \ref{section:results} we describe our main results and conclude in Section \ref{section:discussion}.

Throughout the paper, we adopt the following cosmology:  $\Omega_m$ = 0.3,  $\Omega_{\Lambda}$ = 0.7 and all the relevant quantities are rescaled to h = $H_0/100 \,km\, s^{-1}\, Mpc^{-1}$

\section{The data}\label{section:data}

In this Section we describe the source galaxy and cosmic void catalogs, as well as the morphological information used in this work.

\subsection{The SDSS void catalogue}\label{sdss}

All the galaxies used in this study are drawn from the New York
University Value-Added Galaxy
Catalog\footnote{\url{http://sdss.physics.nyu.edu/vagc/}} (NYU-VACG;
\citealt{Blanton05}), based on 
SDSS/DR7\footnote{\url{http://cas.sdss.org/astrodr7/en}}.
We used the SFR and stellar mass estimates from the MPA
catalogue\footnote{\url{http://www.mpa-garching.mpg.de/SDSS/DR7/}}
\citep{Brinchmann04}. 

As for the void catalog, we use the one presented in \citet{Varela12}. Voids are identified in the galaxy distribution, using a volume limited sample, complete down to magnitude
$M_r-5logh=-20.17$ in the redshift range: $0.01 \leq z \leq 0.12$. Voids are then defined as spherical regions devoid of galaxies.
This  parent catalog includes 630
voids, which, by definition, can host only galaxies fainter than
$M_r-5log(h)=-20.17$.   Void radii range between 10 and 18 $\, h^{-1}\,  Mpc$. In the following, we refer to void radius and void size interchangeably.

In the original catalog, the masked areas of the surveyed region were not
taken into account. These holes in the observed area could give rise to
spurious detection of voids. 
In order to correct for this potential bias we have cross-correlated our void catalog with the SDSS-DR7 coverage mask 
generated using {\it Mangle\footnote{Mangle is a suite of free open-source software designed to deal accurately and efficiently with complex angular masks. Mangle is freely available at \url{http://space.mit.edu/~molly/mangle/}}}. This mask identifies the observed area and masks the {\it holes} due to bad observations, bright stars, satellite trails and other artifacts. More precisely, we determine the geometrical completeness for each void, defined as the fraction of projected area that falls inside the observed sky. 
To compute this completeness, we have randomly distributed 100000 galaxies within  each void and estimated the fraction of galaxies falling outside the visibility region. The larger this number, the higher the probability that the observed void is not a true void. For this reason we apply a threshold to a completeness level of 80\%, below which we exclude the void from the input catalog. This cut reduces the full sample by $\sim10\%$, but ensures a better purity of the void selection. 
Thus, our {\it clean} void catalog contains 566 voids.

\begin{figure}
\includegraphics[width=\columnwidth]{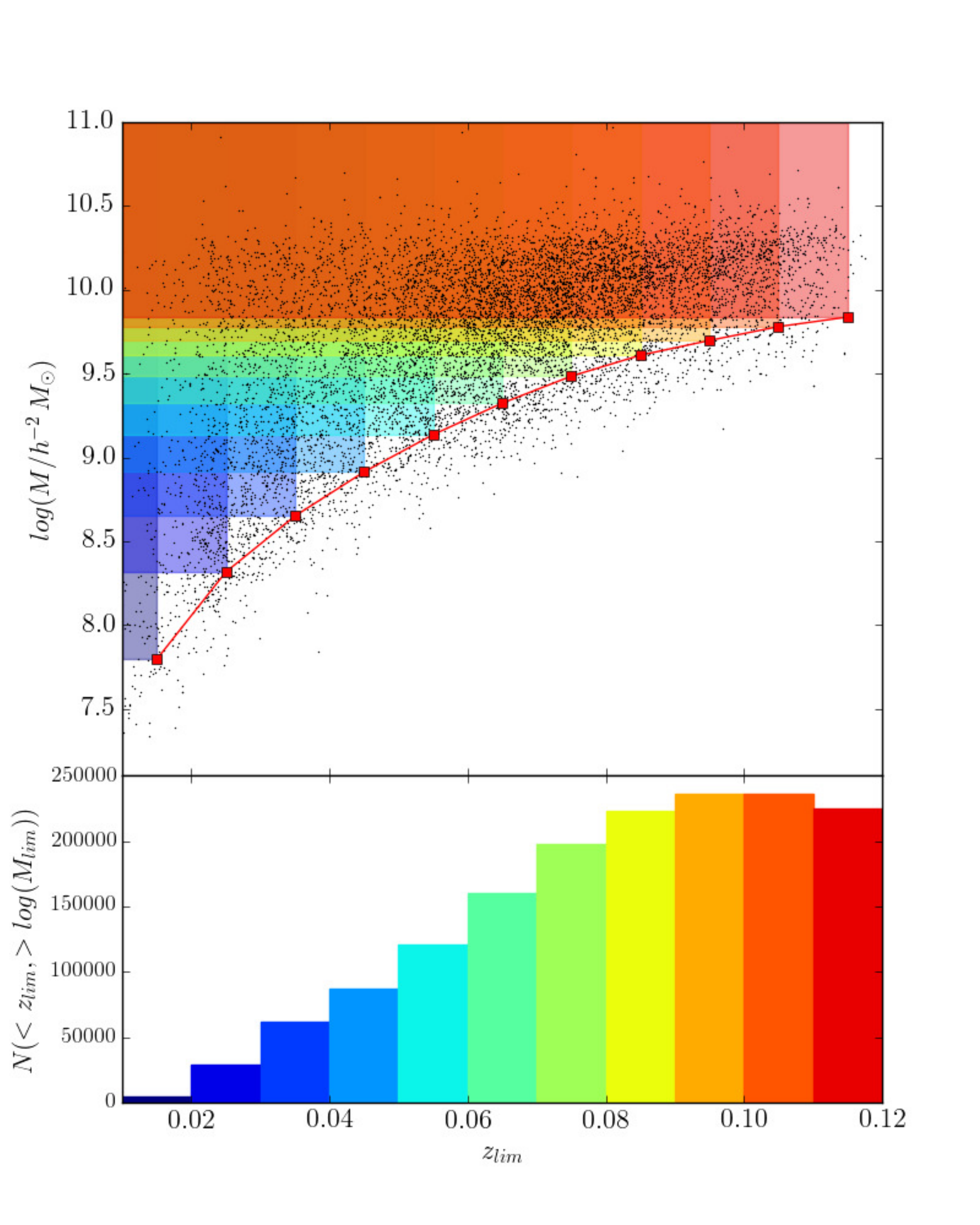}
\caption{Upper panel: galaxies in the parent sample (black points) in the redshift stellar mass
  plane. For the sake of clearness only a randomly selected subsample, including
  10\% of the galaxies, has been plotted. The shaded areas indicate
  the volume-complete samples that we have chosen for the analysis. Lower panel: number of galaxies with redshift less than $z_{lim}$ and
  more massive than the corresponding threshold mass at $z_{lim}$. }
\label{completeness} 
\end{figure}

For consistency with the void catalog, we used for the present analysis only
galaxies fainter than $M_r-5log(h)=-20.17$. Our parent galaxy catalog is constructed by matching  galaxies in  NYU with our void catalog, considering all galaxies with: $d/R_{void} \leq 2.5$, where $d$ is the comoving distance from the center of a void and $R_{void}$ is the void radius. 
It might occur that the same galaxy 
 belongs to the overlapping shells of different voids.  In these
 cases, and unless otherwise stated, we use the multiple occurrences
 of galaxies as different measurements. In Section \ref{subs:morpho} we also consider a sample of void galaxies, defined as those galaxies lying at a distance: $d/R_{void}<1$ from a void.
 In addition, we  build a control sample, 
 including all  galaxies  fainter than
$M_r-5log(h)=-20.17$ and within the same redshift range of the void
sample: $0.01 \le z \le 0.12$. The control sample also includes the galaxies in and around the voids,

Figure \ref{completeness} shows how the galaxies in the parent catalog populate the
redshift - stellar mass plane. For each redshift bin we compute 
a stellar mass threshold, above which the sample can
be considered complete. We chose this mass threshold as the mass above 
which we have 90 \% of the galaxies at that redshift bin.
The colored area shows the volume-limited samples that we have used for
our analysis. The use of volume-limited samples, as defined here, ensures to be immune by the Malmquist bias, when comparing galaxies in voids of different radii (see Section \ref{section:segregation}). In fact,  the largest voids are preferentially located at higher redshift, because the comoving volume spanned by the survey is larger, and thus, when the parent sample is concerned, they contain only bright galaxies. This is not the case when considering samples limited both in redshift and in magnitude.
\subsection{Galaxy Zoo Morphology}

In order to obtain the morphological information, we rely on the
Galaxy Zoo visual morphology. Galaxy Zoo\footnote{\url{http://www.galaxyzoo.org}} \citep{Lintott08} is a citizen science project,
 that provides the visual classification of nearly one million of galaxies.
Each galaxy has been classified by on average 34 users, who could label
galaxies into 'Elliptical', 'Spiral', 'Merger' and 'Don't know'.  The
proportion of classifications in each class has been translated into
raw likelihood. For our analysis we are only interested
in the elliptical and spiral classes. 
We use as a morphological classification, the de-biased
likelihood, that have been corrected for the bias with respect
luminosity, size and redshift, as faint and small size objects tend
to be classified as ellipticals.  
By matching our parent catalog with  Galaxy Zoo  we end up
with 389285  measurement in the parent catalog (6000  of these are void galaxies) and 194446 galaxies
in the control sample.  

\section{Results}\label{section:results}

\subsection{The morphology of void galaxies} \label{subs:morpho}

\begin{figure}
\includegraphics[width=\columnwidth]{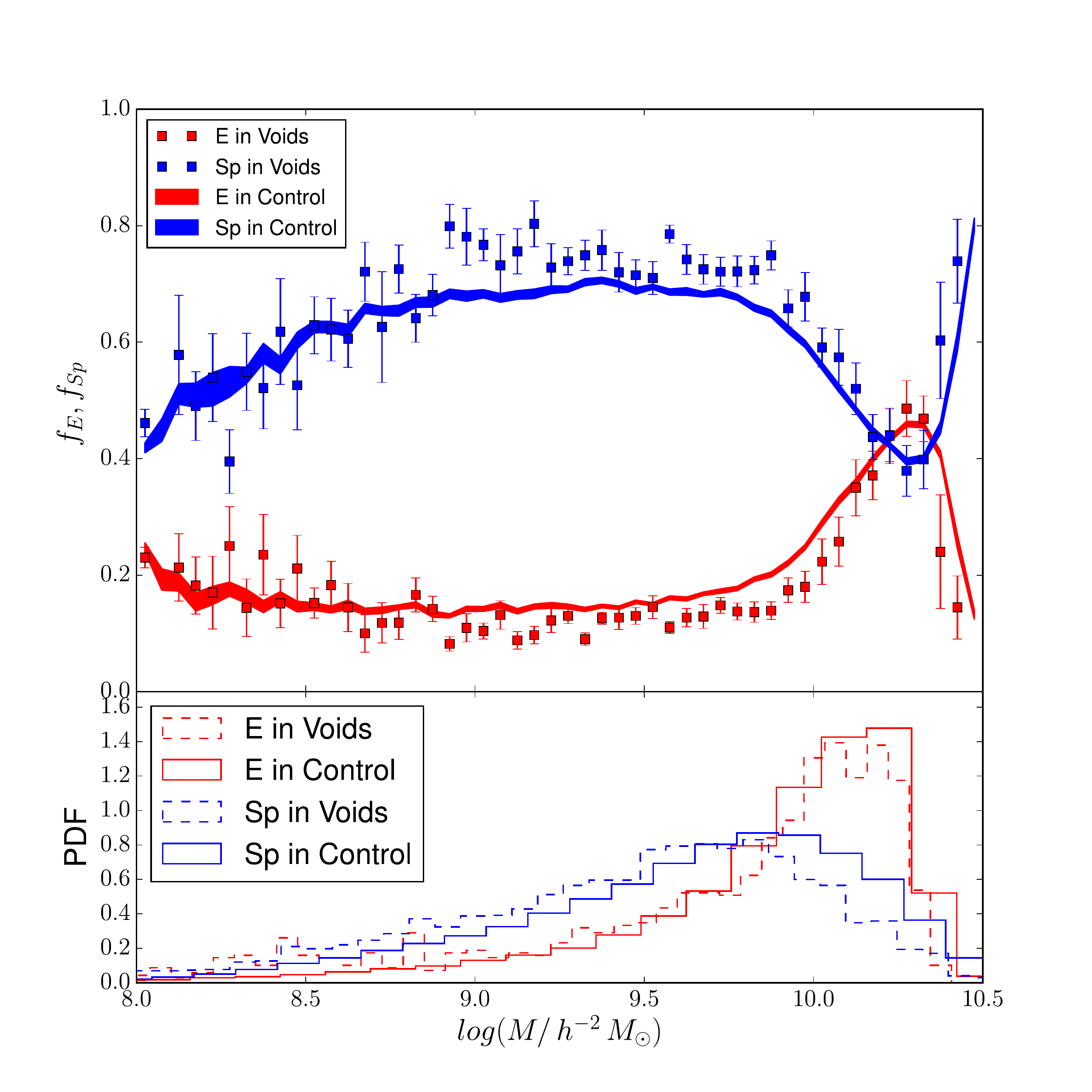}\\
\includegraphics[width=\columnwidth]{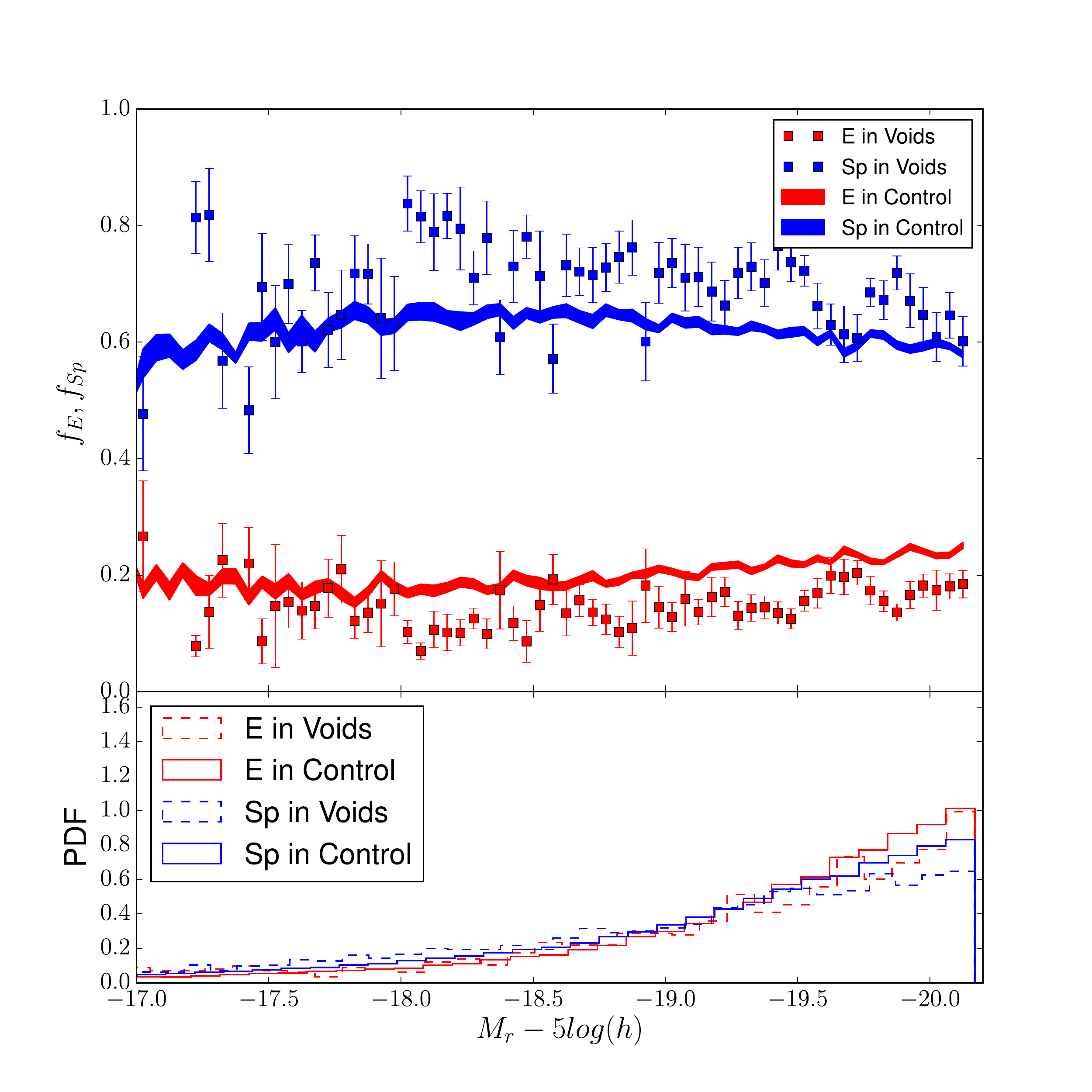}
\caption{Top panel: elliptical (in red) and spiral (in blue)
  fraction as a function of stellar mass for void galaxies
(squares) and for galaxies in the control sample (hatched region). Morphological
fraction are defined from the de-biased  likelihood, as said in the
text. Errorbars are computed from 100 bootstrap resamplings. Bottom inset:
Stellar mass distributions of elliptical (in red) and spiral (in blue) in voids (dashed lines)
and in the control sample (solid lines).  
Bottom panel: ellipticals (in red) and spirals (in blue)
  fraction as a function of absolute magnitude. Bottom inset: Absolute magnitude distributions of elliptical (in red) and spiral (in blue) in voids (dashed lines)
and in the control sample (solid lines). }
\label{morpho_mass}
\end{figure}

One way to discriminate the effect of environment on the evolution of galaxies
is to compare galaxy properties across the same stellar mass or luminosity, in order to isolate pure environmental effects, from the mass bias, i.e. the fact that the low-mass galaxies dominate the low-density environments. 
In Fig. \ref{morpho_mass} we show the fraction of ellipticals (spirals)
as a function of stellar mass and absolute magnitude for void galaxies and for the control sample. At each stellar mass (or absolute magnitude) bin, $f_E$ ($f_{Sp}$) is given by the median de-biased elliptical (spiral) likelihood and the errorbars are computed with 100 bootstrap resamplings. 
At low stellar mass the fraction of ellipticals and spirals is almost constant with stellar mass, whereas for large masses the fraction of ellipticals shows an increase with stellar mass, that gets reflected in a steady decline in the fraction of spiral galaxies. At stellar mass larger than $log(M\,h^2)>10.2$ we see a drop in the elliptical fraction and an increment in the spiral fraction.
This behaviour is due to a sharp selection cut in absolute magnitude that does not translate in a sharp selection in stellar mass, because of the spread in the mass-to-light ratios. The high fraction of massive spirals are indeed edge-on disks, which are likely to be highly extinguished. Their absolute magnitude is thus underestimated and they enter the selection  cut.  
 On the other hand, the morphology shows a linear behaviour with the absolute magnitude, with the fraction of ellipticals (spirals) increasing (decreasing) with luminosity.
At all stellar masses and absolute magnitudes, the fraction of elliptical (spiral) galaxies in
voids is smaller (larger) than in the control sample. We thus see that the void environment shows a pure environmental effect on galaxy evolution, which is independent on the mass bias.
This result is in agreement with the higher fraction of star forming galaxies observed in voids \citep{Ricciardelli14b}.
\subsection{Morphological segregation in voids of different size}\label{section:segregation}
\begin{figure}
\includegraphics[width=1.\columnwidth]{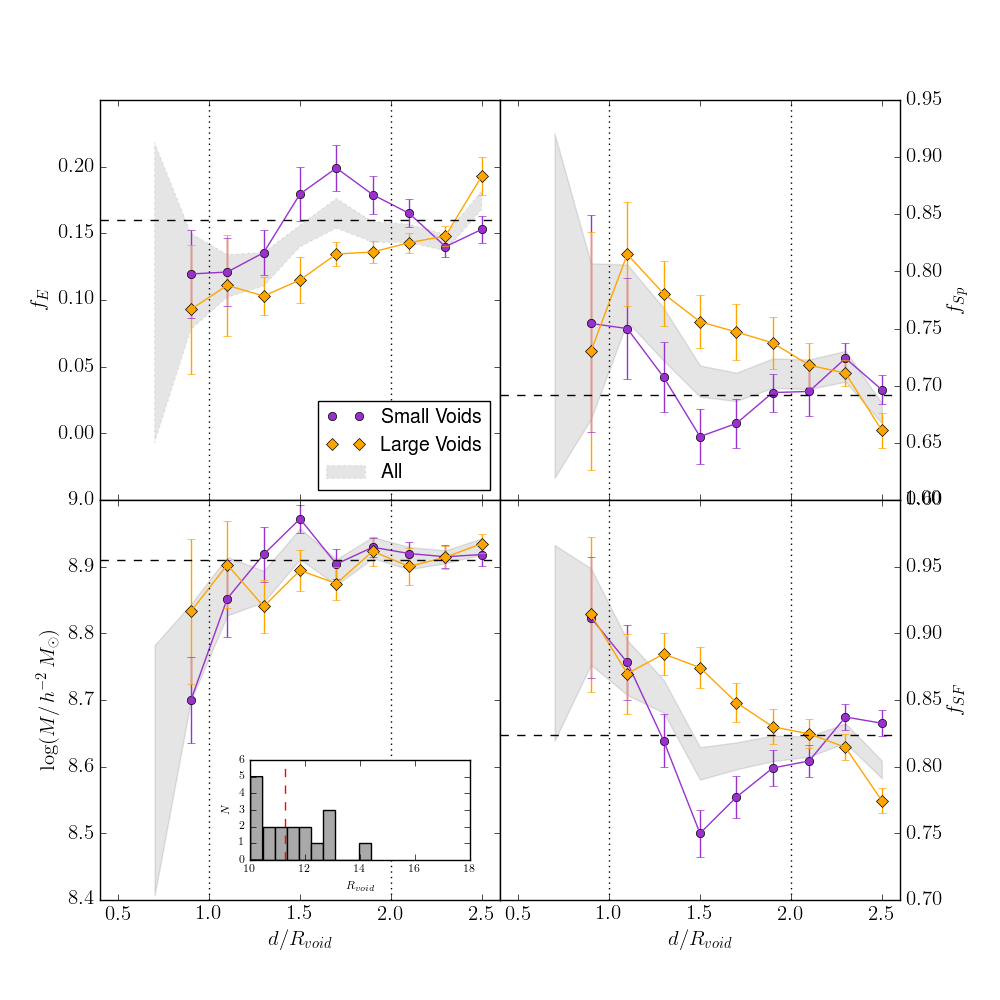} 
\caption{Dependence of galaxy properties, namely the elliptical fraction, spiral fraction, stellar mass and star forming fraction, on the void-centric distance, for voids of different sizes and for the volume-limited sample at $z_{lim}=0.025$, $M_{lim}=10^{8.3}$. The grey shaded area indicates galaxies in and around voids of all sizes, orange symbols indicate galaxies in large voids and purple symbols in small voids. The dashed horizontal line indicates the median value of the control sample. The vertical dashed lines indicate the void edge at $d/R_{void}=1$ and the region of influence of voids, defined as $d/R_{void}=2$.  The inset panel shows the distribution of void radii for this galaxy sample, the median void radius is indicated by the red dashed line. }
\label{morpho_dist_zlow}
\end{figure}
\begin{figure}
\includegraphics[width=1.\columnwidth]{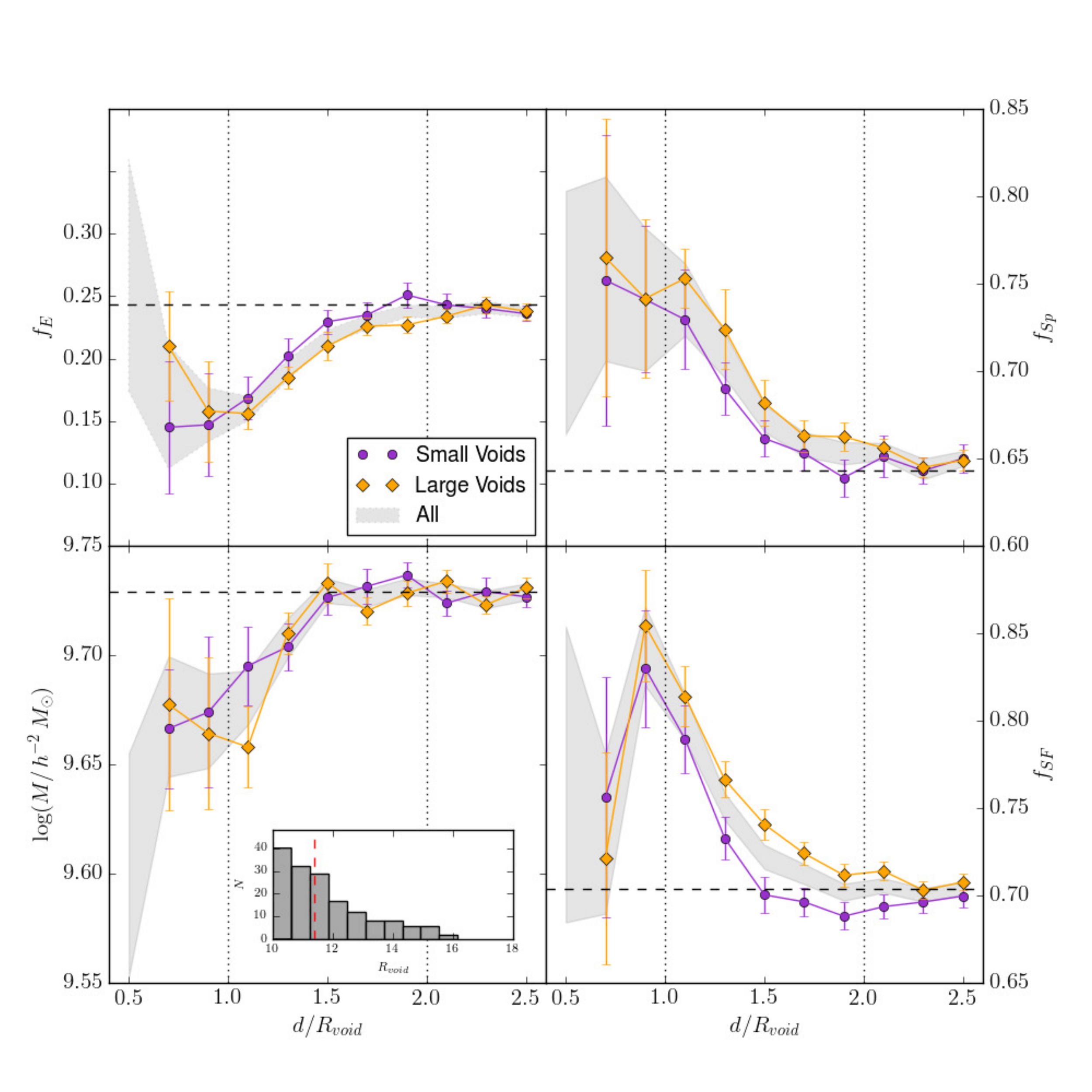}
\caption{The same of Figure \ref{morpho_dist_zlow} for the volume-limited sample: $z_{lim}=0.065$, $M_{lim}=10^{9.3}$.}
\label{morpho_dist_zhigh}
\end{figure}
In this section we explore how galaxy properties vary as a function of void-centric distance and
void size.
For each volume-limited sample, we have divided the void sample in small and large voids, using as a discriminant size the median void radius for that sample, which is in the range 11.3-11.5 $\, h^{-1}\,  Mpc$.
In Figure \ref{morpho_dist_zlow} we show the morphological fractions  as a function of void-centric distance, for the sample with $z_{lim}=0.025$, $M_{lim}=10^{8.3}$. As a comparison, we also show the stellar mass and the fraction of star forming galaxies, $f_{SF}$. The latter is defined as in \citealt{Ricciardelli14b}. Thus, we define as  star forming galaxies, all the galaxies lying above the line separating star forming and passive galaxies on the SFR-stellar mass plane (see their equation 4). For each void-centric bin we present the median value of each property. 
The confidence intervals include two sources of errors. The first is the statistical error computed by means of 100 bootstrap resampling. The second source of error comes from distance uncertainties due to the galaxy peculiar velocities. We assume a normal  distribution for peculiar velocities along the line of sight, whose dispersion is $511/\sqrt(3)\,km\,s^{-1}$ \citep{Agarwal13}. Thus, for each galaxy we pick up a random peculiar velocity from this distribution and shift its distance accordingly. We then compute the standard deviation for 50 Montecarlo simulations. The two errors are then summed up in quadrature to give the confidence intervals shown in Figure \ref{morpho_dist_zhigh}. 

We find a significant correlation between average galaxy properties and void-centric distance, that we refer to as segregation.
The most massive and early-type galaxies are found at large void-centric distances, whereas late-type and star forming galaxies are located at small void-centric distances. 
As a comparison, we show the median value of the control sample as a horizontal dashed line. The properties of  galaxies in the proximity of voids converge to the value of the control sample at void-centric distances $\sim1.5\, R_{void}$, as pointed out in \citealt{Ricciardelli14b}. However, as we say below, the region of influence of voids extends beyond this scale.

The most striking result of Figure \ref{morpho_dist_zlow} is the significant morphological difference in the surroundings of voids of different sizes.
The surroundings of small voids host an excess of early-type galaxies and a lack of late-type galaxies, with respect to their smaller counterpart. Conversely, large voids are preferentially surrounded by late-type galaxies. The trend in the fraction of star forming galaxies is in agreement with that for the late-type galaxies. The trend is particularly significant in the region outside the void, up to void-centric distance $d/R_{void} \sim 2$.  We thus define the region of influence of voids as the region with void-centric distance: $0<d/R_{void}<2$.
Interestingly, the stellar mass does not show any significant dependence on the void size. 
 In Figure \ref{morpho_dist_zhigh} we show the segregation analysis for a sample at higher redshift, namely: $z_{lim}=0.065$, $M_{lim}=10^{9.3}$.
In this sample, as in all the other, the impact of the size of the void on galaxy properties is evident. The strength of the signal is however lower, with respect to the sample at lower redshift. 

\begin{figure}
\includegraphics[width=1.\columnwidth]{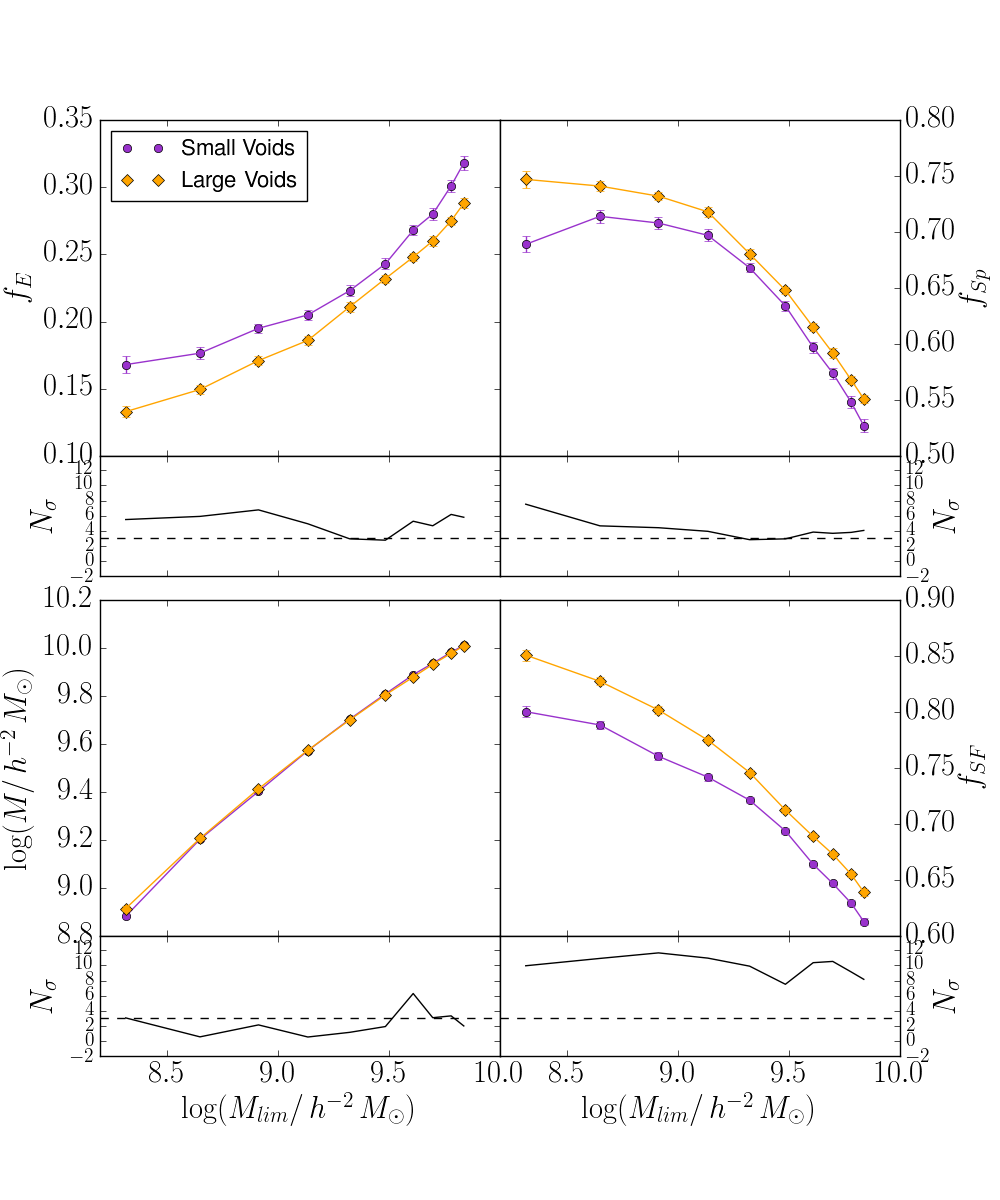}
\caption{Elliptical fraction (upper left panel), spiral fraction (upper right panel), stellar mass (bottom left panel) and star forming fraction (bottom right panel) in small (purple) and large (orange) voids, as a function of the limit mass of the sample, $log(M_{lim})$ (see Figure 1).
The small insets on the bottom of each panel represent the significance of the difference between small and large voids, in terms of $N_{\sigma}$. The $3\,\sigma$ level is highlighted by a dashed line.}
\label{significance}
\end{figure}

To better quantify the significance of the signal, we show in Figure \ref{significance} the median morphological fractions, stellar masses and star forming fractions in the region of influence of voids of different sizes. 
We consider all the volume-limited samples between $z_{lim}=0.025$ ($M_{lim}=10^{8.3}$) and $z_{lim}=0.115$ ($M_{lim}=10^{9.8}$). 
In the lower panels, we show the significance of the signal, that we define as:
\begin{equation}
 N_{\sigma}=\frac{X_{small}-X_{big}}{max[err(X_{small}), err(X_{big})]}\,,
\end{equation} 
where $X$ denotes the galaxy property under consideration ($f_E$, $f_{Sp}$, $log(M)$ or $f_{SF}$). The difference observed between small and large voids is more significant than $3\,\sigma$ for all the samples, whereas no significant difference in stellar mass can be noticed. 
We also note a slight dependence of the signal on the sample, with samples at lower redshift and lower stellar mass limits having the higher strength and larger significance, despite the fact that they are the less populated samples, and thus have the largest statistical error. The dependence on the sample is particularly noticeable in the fraction of spirals.
\section{Discussion and conclusions}\label{section:discussion}

In this paper we explore the morphological properties of void galaxies, using a void catalog from SDSS-DR7. We find that void galaxies have morphology of later type than galaxies in the control sample, even when compared across the same stellar mass or absolute magnitude. Thus, a pure environmental impact on galaxies, beside the mass bias, is present and it is consistent with void galaxies being formed at a later time than galaxies living in more dense environments.

We find a significant correlation between galaxy properties and void-centric distance, that we refer to as segregation,  with less massive, more star forming and later type galaxies being closer to the void center. 
The segregation in stellar mass and star forming fraction was already pointed out by \citet{Ricciardelli14b} and it is here confirmed by the morphological fractions.
The origin of such correlation can be linked to the shape of void density profiles, whose density shows an exponential increases when approaching the void edge \citep{Colberg05, Ricciardelli13, Ricciardelli14a}. Thus, galaxies that are more massive and passive are found towards the edge of the void.
Our result is also consistent with the fact that galaxies in the proximity of filaments show a reversal trend, with more massive and passive galaxies being found closer to the filaments \citep{Malavasi17, Kuutma17}.

The most intriguing result of the paper is that such a segregation depends on the size of the voids.  By splitting the void sample in small and large voids, whose radius is smaller and larger than the median void radius, respectively, we find that late-type galaxies tend to be found in the proximity of voids of large size. The region of influence of voids extends up to $2\,R_{void}$.
We quantify the significance of this difference by comparing the global morphological fractions in the region of influence  of small and large voids. The trend is more significant than $3\,\sigma$ in all the samples analysed. 
We also notice a weak dependence of the significance of the result on the sample,  with samples at low redshift and low stellar mass displaying the more significant signal. Such a dependence could be due to an enhanced sensitivity of the dwarf galaxies to the large-scale environment.
The trend in morphology is confirmed by the star forming fractions, that are more numerous in and around voids of small size.
Conversely, the stellar mass does not show any significant dependence on the size of the void.  Therefore, the different morphological and star formation properties in voids of different size can be interpreted as due to pure environmental effects.

In this work we show, for the first time, that the size of a void can have an impact on the evolutionary history of the galaxies living in its surroundings. 
A possible interpretation for this effect can lie in the dichotomy in the void population introduced by \citet{Sheth04}. Voids embedded in large-scale underdensities, {\it void-in-void}, grow in time by merging of small-scale voids. On the other hand, voids that are embedded in large scale overdensities, {\it void-in-cloud}, shrink and could eventually disappear as a consequence of the collapse of the overdense patch. These different evolutionary paths among voids of different sizes can also explain the overcompensated shells observed in small voids \citep{Ceccarelli13, Paz13, Hamaus14}. It is worth to notice however, that when voids are defined according to a dynamical criterion as expanding domains, voids of different size do not show significant different properties \citep{Ricciardelli13, Ricciardelli14a}. 
Therefore, our results could also depend on the void definition adopted. We plan to test this in a future work.

\acknowledgments

ER acknowledges support from a MHV grant from the SNSF.  The work of AC is supported by the STARFORM Sinergia Project funded by the SNSF.
We thank the referee for his/her useful comments that help improving the manuscript.


\end{document}